\documentclass[aps,prl,twocolumn,a4paper,10pt,notitlepage,footinbib,superscriptaddress,longbibliography]{revtex4-2}
\usepackage[normalem]{ulem}
\usepackage[english]{babel}
\usepackage{lmodern}
\usepackage[latin9]{inputenc}
\usepackage{endnotes}
\usepackage{amssymb,amsmath,amsfonts}
\usepackage{textcomp}
\usepackage{braket}
\usepackage{ulem}
\usepackage{soul}
\usepackage{comment}
\usepackage{graphicx}
\usepackage{dcolumn}
\usepackage{bm}
\usepackage{xcolor}
\usepackage{soul}

\def\r{\bm{r}}
\def\p{\bm{p}}

\usepackage{xcolor}
\definecolor{blueprl}{RGB}{13.0, 18.0, 180.0 }
\usepackage[colorlinks=true, allcolors=blueprl]{hyperref}

\usepackage{orcidlink}

\begin{document}

\title{Phase Separation Kinetics in a Polar Active Field Model}

\date{\today}
\author{Massimiliano Semeraro~\orcidlink{0000-0001-8273-4232}}
\email{massimiliano.semeraro@uniba.it}
\affiliation{Dipartimento  Interateneo di  Fisica,  Universit\`a  degli  Studi  di  Bari, via  Amendola  173,  Bari,  I-70126,  Italy}
\affiliation{INFN, Sezione  di  Bari,  via  Amendola  173,  Bari,  I-70126,  Italy}

\author{Leticia F. Cugliandolo~\orcidlink{0000-0002-4986-8164}}
\affiliation{Sorbonne Universit\'e, Laboratoire de Physique Th\'eorique et Hautes Energies, CNRS-UMR 7589, 4 Place Jussieu, 75252 Paris Cedex 05, France}

\author{Giuseppe Gonnella~\orcidlink{0000-0002-1829-4743}}
\affiliation{Dipartimento  Interateneo di  Fisica,  Universit\`a  degli  Studi  di  Bari, via  Amendola  173,  Bari,  I-70126,  Italy}
\affiliation{INFN, Sezione  di  Bari,  via  Amendola  173,  Bari,  I-70126,  Italy}

\author{Adriano Tiribocchi~\orcidlink{0000-0002-5314-9664}}
\email{adriano.tiribocchi@cnr.it}
\affiliation{Istituto per le Applicazioni del Calcolo, Consiglio Nazionale delle Ricerche, Via dei Taurini 19, Rome, I-00185, Italy}
\affiliation{INFN Tor Vergata, Via della ricerca scientifica 1, Rome, I-00133, Italy}

\begin{abstract}
A milestone of phase separation kinetics is the emergence of universal power laws $\sim t^{1/z}$ governing the domain growth evolution. We investigate a phase-separating polar active model comprising a scalar density field with an advective coupling to a polarization field. Our analysis reveals a novel $\sim t^{0.6}$ regime, which agrees well with the accelerated growth recently observed in simulations of polar active particles. We provide analytical arguments to explain how advection facilitates the creation of topological defects and compresses the domains leading to faster growth. 
We also show that the $\sim t^{0.6}$ regime is robust to several model generalizations.
\end{abstract}
\maketitle

Phase separation kinetics holds a special place in non-equilibrium physics, as it is characterized by general properties such as dynamical scaling and simply classifiable domain growth laws \cite{hohenberg1977, bray1994}.  The  domain average  size typically grows as  $ L(t) \sim t^{1/z}$ with the dynamic exponent $z$ assuming few integer universal values which depend on the physical mechanism acting during phase separation. For example, in binary mixtures $z=3$ for the diffusive case or $z=2$ when the order parameter is not  conserved, while $z$ takes two other values when hydrodynamics comes into play, depending on the relevance of the viscous or the inertial term in the momentum conservation equation~\cite{bray1994, onuki2002}.  The presence of additional driving forces beyond the thermodynamic ones that describe the relaxation towards equilibrium makes this phenomenology richer, yet still with  general features. Examples are phase separating systems under shear~\cite{onuki1995, corberi1998, stansell2006} or active systems~\cite{marchetti2013, cates2015, elgeti2015, bechinger2016, gompper2020}.

In this Letter, we study the aggregation process in a two-dimensional model of polar matter under self propulsion. Self-propulsive forces are at the origin of many interesting phenomena found in active matter systems. In the case of active Brownian particles, despite the absence of attractive interactions, a motility-induced phase separation is observed between a low-density gas-like phase and dense stable aggregates~\cite{tailleur2008, fily2012, buttinoni2013, digregorio2018, elsen2018, klmaser2019, paoluzzi2022, cates2025}. The related phase separation kinetics has been studied, e.g., in \cite{stenhammar2013, speck2014, redner2016, caporusso2020}, for particle and continuous models, finding the diffusive value $z\sim 3$ for the cluster size growth. Since many active systems are composed of intrinsically polar units with head-tail direction~\cite{toner2005, marchetti2013, needleman2017, chate2020, borthne2020, paoluzzi2024}, such as motile cells \cite{fodor2018, laang2024}, active gels and filaments \cite{schaller2010, vliegenthart2020}, bacteria \cite{lopez2015} or synthetic self-propelled rods \cite{yang2010, bar2020}, it is natural to consider models catching the ability to move persistently in a locally preferred direction. The choice of local  orientation can be induced by explicit alignment interactions between close particles, as in the case of flock models~\cite{vicsek1995, toner1995, cavagna2014, solon2015, negi2022}, or can be an intrinsic property of the dynamical units related to their shape, as in the case of active rods~\cite{ginelli2010} or active Brownian dumbbells~\cite{linek2012, suma2014, cugliandolo2017, clopes2022}. For the latter system, the growth law $L(t)\sim t^{0.6}$ was measured with molecular dynamic simulations~\cite{caporusso2024}. One purpose of this work is to propose a field theory that captures this accelerated growth and provides an interpretation of its origin. More generally, we aim to understand how self-propulsion affects phase separation in polar systems.

We consider a two-dimensional model comprising a scalar field $\phi$, representing the local density, and a polar field $\p$, describing the local orientation of the active component. The system undergoes phase separation, with polarization slaved to the dense phase. Similar approaches were adopted to study emulsions and droplets of liquid crystals~\cite{cates2018} or mixtures of passive and active phases~\cite{giomi2008, bonelli2019}. From the numerical study of the field model, we will reveal a growth regime of the polar domains characterized by the growth $\sim t^{0.6}$ for sufficiently intense self-propulsion strength. We will discuss the relevance of compression forces due to self-propulsion and of topological defects~\cite{vafa2022, angheluta2025} for the aggregation process. Using scaling arguments, we will relate the  value of the growth exponent to the size dependent compressibility properties of the dense droplets, finding that  $1/z\sim 3/5 =0.6$.

The equilibrium properties of our mixture are encoded in the following free energy functional
\begin{eqnarray}
&& F[\phi,{\p}]=\int d^2x \; \Big\{\frac{\alpha_\phi}{4\phi_{cr}}\phi^2(\phi-\phi_0)^2+\frac{k_\phi}{2}|\nabla \phi|^2
\nonumber\\
&& \quad\qquad\quad -\frac{\alpha_{\p}}{2}\frac{\phi-\phi_{cr}}{\phi_{cr}}|{\p}|^2
+\frac{\alpha_{\p}}{4}|{\p}|^4 +\frac{k_{\p}}{2}(\nabla{\p})^2\Big\}
\, ,
\;\;\;\;
\label{eq:free_en}
\end{eqnarray}
where $\phi$ is a density and $\p$ a polarization field, and dependencies on $\bm{r}$ and $t$ are implicit. The first term guarantees the existence of two coexisting minima, $\phi=0$ and $\phi=\phi_0$, which, in absence of polarization, set the equilibrium values of the two phases of $\phi$. The second term is an energetic penalty associated to the fluid interfaces. The parameter $k_\phi$ controls the surface tension and interface width, which are equal to  $\sigma=\sqrt{8\alpha_\phi k_\phi/9}$ and $\xi=\sqrt{2k_\phi/\alpha_\phi}$, respectively. The third and fourth terms set the bulk properties of the polar field.  Their minima are at $|\p| = \sqrt{(\phi-\phi_{cr})/\phi_{cr}}$ when $\phi>\phi_{cr}$ and $0$ otherwise. The polarization is confined to the dense phase and  at $\phi_{cr}=\phi_0/2$ there is an isotropic-to-polar transition. Finally, the last term $(\nabla\p)^2\equiv\sum_{ij}(\partial p_i/\partial {r}_{j})^2$ ($i,j=1,2$ dimensional indices) accounts for spatially inhomogeneous deformations and corresponds to the single elastic constant approximation~\cite{degennes1993}. The dynamics are governed by the following equations
\begin{eqnarray}
\label{eq:phi}
\partial_t{\phi}+\lambda \nabla \cdot({\p}\phi) &=& M\nabla^2\mu_\phi,
\\
\label{eq:P}
\partial_t{\p}&=& -\Gamma\bm{\mu}_{\p},
\end{eqnarray}
where $\lambda$ is a positive constant gauging the strength of the advective term, $M$ is the mobility, $\Gamma$ is the rotational viscosity, $\mu_{\phi}=\delta F/\delta \phi$ and $\bm{\mu}_{\p}=\delta F/\delta{\p}$ are the chemical potential and molecular field, respectively. The polarization acts as a self-propulsion velocity for the scalar density. This model can be regarded as a simplified version of the Toner-Tu model for dry active matter~\cite{toner1995, marchetti2013, worlitzer2021} to which an advection contribution proportional to $(\p\cdot\nabla) \p$ could be also included (we will shortly discuss its effects below and in Supplemental Material (SM) Sec.~IX \cite{SM}).

We solved Eqs.~(\ref{eq:phi})-(\ref{eq:P}) by means of a finite difference scheme with periodic boundary conditions on square lattices with linear sizes ranging from $N=128$ to $N=1024$. If not stated otherwise, we fixed the parameters as follows: $\phi_0=2$ (thus $\phi_{cr}=1$), $M=1$, $\alpha_{\phi}=0.1$ and $k_{\phi}=0.5$ (hence $\sigma\sim 0.21$ and $\xi\sim 3.16$), $\Gamma=1$, $\alpha_{\p}=0.1$ and $k_{\p}=0.04$.  Also, we fixed the lattice spacing $\Delta N=1$ and time-step $\Delta t=10^{-2}$ (further details and mapping to physical units in SM Sec.~II). We studied the phase separation kinetics for different values of the advection strength $\lambda$ starting from an initial mixed state, where $\phi$ is randomly distributed in the range $[0.9, 1.1]$ while $\p$ has unitary modulus and random orientation. 

\begin{figure}[t!]
\centering 
\includegraphics[width=0.99\columnwidth]{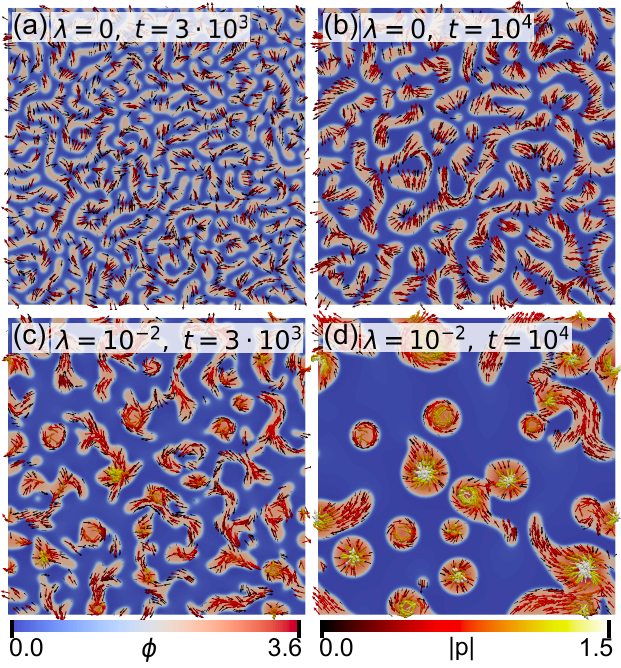}
\caption{\footnotesize{
Typical snapshots ($N=512$) at different times with $\lambda=0$~(a)-(b) and $\lambda=10^{-2}$~(c)-(d). The density field and modulus of the polarization are colored according to the codes in the bars. }} 
\label{fig:fig1}
\end{figure}

{\it Dynamic domain structure.}
In Fig.~\ref{fig:fig1} we show system configurations at two times after initialization and for two values of $\lambda$ (see also Movie S1). 
In the absence of advection, after an initial nucleation regime [Fig.~\ref{fig:fig1}(a)], domains of the polar phase grow with time and form elongated structures [Fig.~\ref{fig:fig1}(b)]. Note that the lack of a predefined orientation of the polar field at the fluid interface favors the formation of defect-free domains with uniform polarization. At later times (see SM Sec.~IV), the morphology does not substantially change until only a few domains survive.

Instead, sufficiently strong advection significantly affects the morphology (Movie S2). More rounded shapes appear [Fig.~\ref{fig:fig1}(c)] and quite regular droplets characterize the late stages of phase segregation [Fig.~\ref{fig:fig1}(d)]. As clear in Fig.~\ref{fig:fig1}(d), the polarization field organizes in configurations with integer-charge defects, which can be vortices, inward-pointing asters, or intermediate spiral configurations, as defined in SM Sec.~III and also shown in Secs.~II and IV for additional cases (similar polarization structures were measured in molecular dynamic simulations of active dumbbells, see Fig.~8 in~\cite{Petrelli18}). One can also observe more complex structures resulting from the merging of two (or more) domains and containing multiple defects.
 
\begin{figure}[t!]
\centering 
\includegraphics[width=0.99\columnwidth]{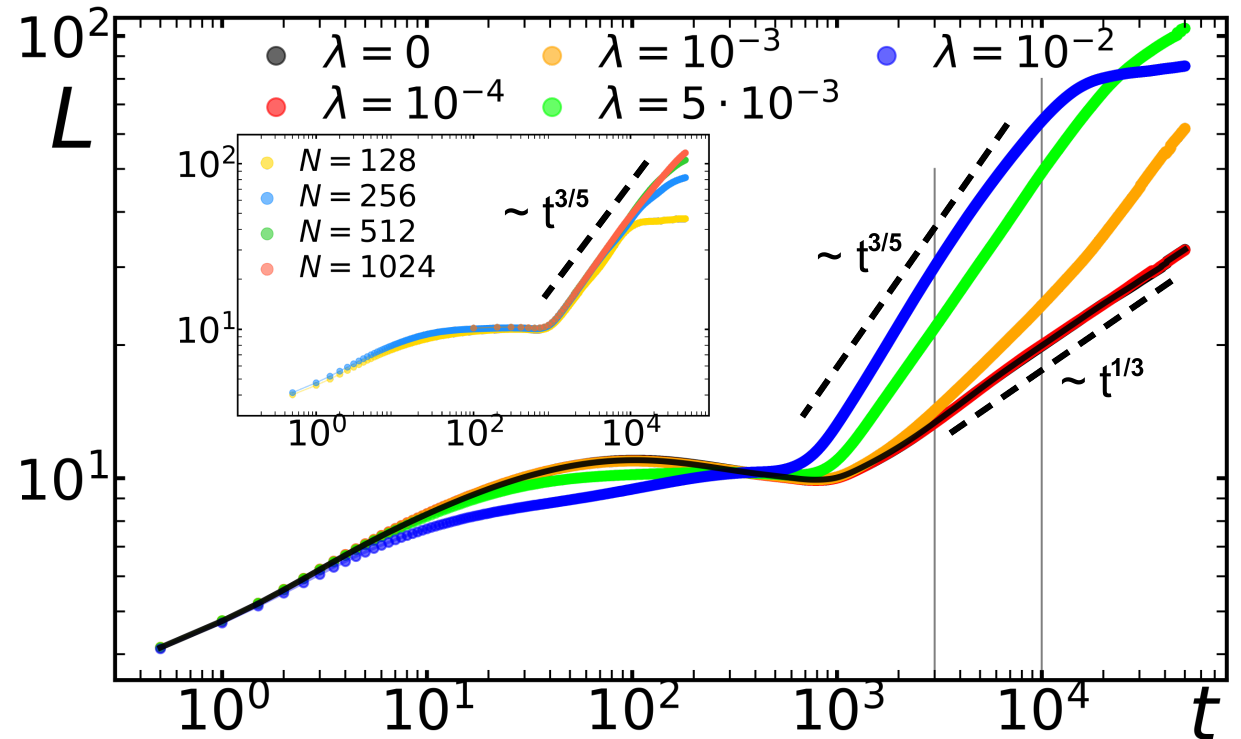}
\caption{\footnotesize{The time dependence of the typical length $L$ for different values of $\lambda$ ($N=512$). Each curve has been averaged over $5$ independent runs. The dashed black segments show the $\sim t^{1/3}$ and $\sim t^{3/5}$ behaviors. The vertical thin black lines indicate the instants at which the snapshots in Fig.~\ref{fig:fig1} were taken. The inset reports the growing length for $\lambda=5\cdot 10^{-3}$ for different system sizes. }}
\label{fig:fig2}
\end{figure}

{\it The growing length.}
The morphological differences arising from varying advection strength $\lambda$ are reflected in the growth of the average domain size. We measure the typical length scale $L(t)$ as the inverse of the normalized first moment of the density structure factor (see SM Secs.~II and VI). Fig.~\ref{fig:fig2} shows the evolution of $L(t)$ for different values of $\lambda$. After an initial transient phase, $t \lesssim 10^{-3}$, which is largely independent of $\lambda$, differences begin to emerge. In the absence of advection ($\lambda = 0$) and for very small $\lambda$, we observe the expected diffusive growth with exponent $z \sim 3$. At $\lambda = 10^{-3}$, we find a crossover from diffusive behavior to a faster growth, which remains stable as $\lambda$ is further increased. This regime is characterized by an exponent $1/z \sim 0.6$
, and the larger the value of $\lambda$, the earlier it sets in. As highlighted by the vertical lines in Fig.~\ref{fig:fig2}, the snapshots in Fig.~\ref{fig:fig1} were taken during the algebraic growth phase. At very late times, the curve for $\lambda = 10^{-2}$ in Fig.~\ref{fig:fig2} reaches a plateau. This behavior can be attributed to the formation of large, regular domains, each containing a central topological defect and a symmetric internal polarization structure (see SM Sec.~IV). As a result, advection effects become balanced, and domain growth is almost completely arrested, or proceeds more slowly via diffusion alone, since the remaining domains are far apart (see SM Fig.~S3). The inset shows $L(t)$ for $\lambda = 5 \cdot 10^{-3}$ and various system sizes, demonstrating that arrest occurs later for larger systems.

\begin{figure}[t]
\centering 
\includegraphics[width=0.99\columnwidth]{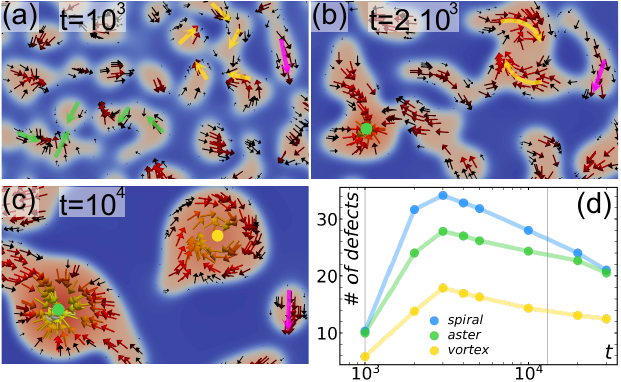}
\caption{
\footnotesize{(a)-(c) Snapshots of an enlarged area of the system with $\lambda = 10^{-2}$ from Fig.~\ref{fig:fig1} at the times indicated in the panels. Green and yellow arrows show the direction of motion of some small clusters which, after colliding, give rise to larger clusters containing an inward aster and a vortex defect, respectively. The magenta arrow indicates the direction of motion of a small isolated domain. Dots mark the locations of the defects (details in SM Sec.~II). The density field and modulus of the polarization are colored as in Fig.~\ref{fig:fig1}. (d) Evolution of the number of defects (averaged over 6 independent runs) for a system with $N = 1024$ and $\lambda = 10^{-2}$. Thin vertical lines indicate the approximate boundaries of the $\sim t^{3/5}$ regime.}}
\label{fig:fig3}
\end{figure}

{\it Topological defects and compression.}
We aim to build an argument to explain why advection induces faster growth of the dense phase. As previously noted, there is a morphological difference between the cases $\lambda = 0$ and $\lambda \neq 0$: in the former, no defects are present, whereas in the latter, defects do appear. A closer examination of domain formation and growth is provided in Fig.~\ref{fig:fig3}. In the early stages, small domains with $\phi\sim\phi_{eq}=1+\sqrt{1+\alpha_{\p}/(2\alpha_\phi)} \sim 2.2$  (see SM Sec.~I) start moving in the direction of their inner uniform polarization [Fig.~\ref{fig:fig3}(a)]. The large colored arrows indicate the direction of motion of a few of these domains, which subsequently collide, merge  [Fig.~\ref{fig:fig3}(b)] and continue to grow [Fig.~\ref{fig:fig3}(c)]. According to the relative polarization orientation of the colliding domains, different topological defects emerge	\cite{kruse2004, elgeti2011, husain2017, mondal2023, popli2025}. A key consequence of this is that, in configurations like asters, the polarization vectors point toward the defect core, compressing the droplets and enforcing a non-uniform density. This effect is evident in Fig.~\ref{fig:fig3}(c), where the concentration field near the defect cores reaches values almost twice that of $\phi_{eq}$.

The time evolution of the number of different types of defects is shown in Fig.~\ref{fig:fig3}(d). The number of all kinds of defects first increases, reaching a simultaneous maximum
within the regime $\sim t^{0.6}$. Since small domains evaporate while larger ones may come into contact and coalesce, the  number of all defects starts decreasing monotonically. At all times, spirals are more frequent than asters, which in turn are more common than vortices. Compression effects are stronger in domains with asters or spirals, therefore we focus on individual droplets with aster configurations. This phenomenology applies to sufficiently large $\lambda$ and system size.

\begin{figure}[t!]
\centering 
\includegraphics[width=0.99\columnwidth]{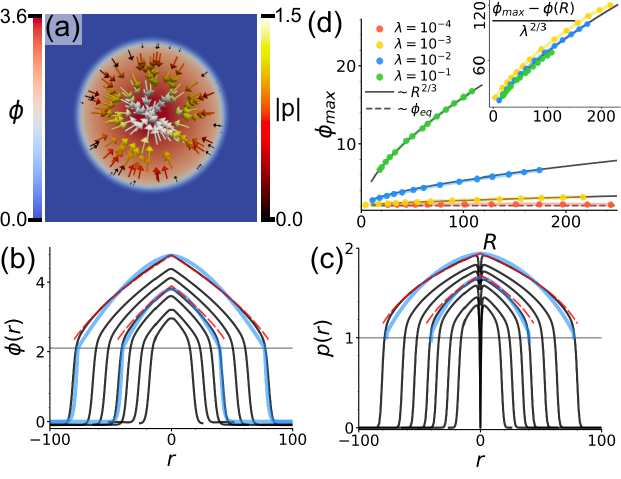}
\caption{\footnotesize{(a) Snapshot of a droplet with an inward-aster configuration ($R \sim 30$, $\lambda=10^{-2}$). (b) and (c) Density and polarization magnitude profiles along diameters with radius $R$ increasing from $\sim 20$ to $\sim 180$. The black lines report the numerical data. The dashed red curves represent the numerical solution of  Eq.~(\ref{eq:phi-r}) for two sample droplets and stop at $r=R$. The blue curves correspond instead to Eq.~(\ref{eq:dens_prof_refine}) for the bulk behavior and approximate the interface with the equilibrium expression for a scalar field theory. (d) $\phi_{max}$ against $R$ for different values of $\lambda$. The continuous and dashed lines highlight the $\sim R^{2/3}$ and $\sim \phi_{eq}$ trends, respectively. The inset shows curve collapse for different $\lambda$ values.}}
\label{fig:fig4}
\end{figure}

{\it Droplet evaporation.}
To understand how advection facilitates the growth of the polar phase, we study  the evolution of a single spherically symmetric droplet with aster polarization, as the one in Fig.~\ref{fig:fig4}(a). We estimate the time dependence of its linear size $R$, which we associate to $L(t)$, from the balance between  the rate of change of its mass and the dominant flow of mass across its surface. A common assumption is that the interior of the droplet is in a quasi-stationary state~\cite{bray1994}. In models where the density profile is uniform, the droplet mass scales as  ${\mathcal M} \propto \phi_0 R^d$ and its time derivative as  $d{\mathcal M}/dt \propto \phi_0 R^{d-1} dR/dt$. Comparing this to the flux of the order parameter through the interface, which scales as $R^{d-1} \nabla \mu_\phi \propto R^{d-1} R^{-2}$ with $\mu_\phi$ denoting the chemical potential, immediately yields  $R \sim t^{1/3}$~\cite{bray1994}. In the advected model, however, the density within the domains is not uniform, and the dominant contribution to the flux may not arise from surface tension. Therefore, in order to apply a similar argument, we first determine the stationary density profile and we then evaluate the main contribution to the flux when $\lambda \neq 0$. 

{\it Stationary profiles in a spherically-symmetric aster droplet.}
In such a domain, both $\phi$ and $\bm{p}$ depend on the radial coordinate $r$. Moreover, the polarization points inwards, $\bm p=-p(r) \hat e_r$. At stationarity, dropping the time derivative in Eq.~(\ref{eq:P}), we have $\alpha_{\bm{p}} p^2 \bm{p} -\alpha_{\bm{p}} (\phi-1) \bm{p} - \kappa_{\bm{p}}  \nabla^2 \bm{p}=0$. Then, neglecting the contribution $\kappa_{\bm{p}}  \nabla^2 \bm{p}$ which is reasonable far from the origin, there are two possibilities: $p(r) = 0$, which is the polarization outside the droplet, or $p(r)=\sqrt{\phi(r)-1}$, the configuration within the droplet. Close to the origin there is a dip in $p$, which is consistent with the location of the defect (see Fig.~\ref{fig:fig4}(c) and SM Fig.~S11).

Concerning the density field, the stationary limit of Eq.~(\ref{eq:phi}), which is analyzed in SM Sec.~VII with a number of suitable approximations, yields
\begin{equation}
\begin{split}
&(4-\phi)\sqrt{\phi-1} - c_\phi {\rm arctan} \sqrt{\phi-1} =\\
 &- \dfrac{\lambda}{2M \alpha_\phi} (R-r) + (4-\phi_{eq}) - c_\phi \arctan\left(\sqrt{\phi_{eq}-1}\right)
\end{split}
\label{eq:phi-r}
\end{equation}
with $c_\phi= 2-\alpha_{\bm{p}}/(2\alpha_\phi)$.
As shown in Fig.~\ref{fig:fig4}(b), the density decays monotonically from its maximum $\phi_{max}$ at $r=0$ to $\phi_{eq}$ at $r=R$, defined as the radius of the droplet. An implicit expression for $\phi_{max}$ as a function of $R$ and $\lambda$ 
follows from Eq.~(\ref{eq:phi-r}) evaluated at the center of the droplet. Arguing that the term scaling as $\phi^{3/2}$ dominates over the arctan one and other constants, we deduce 
\begin{equation}
\phi_{max} \sim [ \lambda/(2M\alpha_\phi) \, R ]^{2/3} 
\label{eq:phimax}
\end{equation}
 for large $R$.
As an alternative approximate expression for the density in the bulk of the droplet we use the Ansatz
\begin{eqnarray}
    \phi(r)=\theta(R-r)\left[(\phi_{eq}-\phi_{max})\left(\frac{r}{R}\right)^c+\phi_{max}\right]
    \label{eq:dens_prof_refine}
\end{eqnarray}
 with $\phi_{max}$ as in Eq.~(\ref{eq:phimax}) (details in SM Sec.~VII). This form is explicit and  simpler to deal with, with $c >1$ the only parameter still to be fixed.

One such typical spherical aster droplet is shown in Fig.~\ref{fig:fig4}(a). The numerical profiles of $\phi$ and $p$ are plotted in Fig.~\ref{fig:fig4}(b) and (c) with black lines for $\lambda=10^{-2}$ and  $R$ ranging from $\sim 20$ to $\sim 180$. The numerical solution of Eq.~(\ref{eq:phi-r}) for $\phi$ and its companion for $p$ are plotted in red for two sample values of $R$, showing very good agreement with the numerics (apart from the dip in $p$, more details in SM Sec.~VII). The expression (\ref{eq:dens_prof_refine}) with  $c=1.5$ is  plotted with blue lines in Fig.~\ref{fig:fig4}(c) and (d) and it also well compares with the numerical measurements not too close to the interface. Finally, the numerical verification of Eq.~(\ref{eq:phimax}) is shown Fig.~\ref{fig:fig4}(d) for several values of $\lambda$. In the inset we also demonstrate the $\lambda^{2/3}$ dependence.

{\it Evolution of the droplet radius.}
We integrate the $\phi$ profile in Eq.~(\ref{eq:dens_prof_refine}) over the spherical volume of a droplet to estimate its total mass
\begin{equation}
{\mathcal M} \propto \phi_{max} R^2 \sim  \left( \dfrac{\lambda}{2M\alpha_\phi} \right)^{2/3} \!\!\! R^{8/3} 
\end{equation}
at large $R$.
For sufficiently large $\lambda$, the mass flux through the droplet boundary is dominated by the advection term and, neglecting the usual term proportional to the mobility, we get:
\begin{eqnarray} 
 && 
 - \lambda 
 \int_S  d^2 x \; \nabla \cdot (\p \phi) 
 =
  - \lambda 
  \int_{\partial S} \! d\ell \; \phi \p \cdot \hat r
  =
  4\pi  \lambda \, R
 \; .
 \end{eqnarray}
The assumption is validated with several numerical tests in SM Sec.~VIII. Finally, comparing the rate of mass change to the mass flux through the boundary, one gets
 \begin{equation}
 \left( \dfrac{\lambda}{M\alpha_\phi} \right)^{2/3} \!\!\!\! R^{5/3} \, \dfrac{dR}{dt} 
\sim 
\lambda \, R
\implies
R\sim c_R t^{3/5}~,
\end{equation}
with $c_R=\lambda^{1/3} (M\alpha_\phi)^{2/3}$. The $\lambda^{1/3}$ dependence is confirmed in SM Fig.~14(d).

{\it Extended models.}
We verified that the growth law $1/z\sim 0.6$ is robust against several model generalizations: the introduction of advection to the polarization field (Movie S3 and S4), the extension of the chemical potential to that of Active Model B \cite{wittkowski2014} (Movie S5 and S6), and the addition of a splay contribution to the free energy which favors aster or vortex configurations (Movie S7 and S8) depending on the sign of the elastic constant. Advective effects consistently outweigh the characteristic dynamics of the corresponding non-polar models (details in SM Sec.~IX).

{\it Conclusions.}
In this Letter we showed numerically, and we justified with scaling arguments, that phase separation in a polar active system is faster, $\sim t^{0.6}$, than in simpler scalar models. We argued that the advective coupling of the scalar to the polar field induces the formation of topological defects in the domains formed by colliding clusters. Its concomitant compression thus facilitates growth. Conversely, when advection is negligible, domain growth relies on diffusion, surface tension, and ripening, which occur on slower time scales. These results demonstrate the importance of advection in determining the growth law in phase separation of mixtures with a polar component. They are in line with the fast growth measured with molecular dynamics of polar active Brownian particles~\cite{caporusso2024}, thus establishing a bridge between field-theoretical predictions and particle-based numerical simulations.

\begin{acknowledgments}
{\it Acknowledgments.} Numerical calculations have been made possible through a Cineca-INFN agreement, providing access to HPC resources at Cineca. M.S. and G.G. acknowledge support from INFN/FIELDTURB project and from MUR projects PRIN 2020/PFCXPE, PRIN 2022 PNRR/P20222B5P9 and Quantum Sensing and Modelling for One-Health (QuaSiModO).
L.F.C acknowledges funding from ANR-20-CE30-0031 THEMA. We thank N. Rana for discussions.
\end{acknowledgments}

\bibliography{Bibliography.bib}

\end{document}